\documentclass[referee]{raa}            

\usepackage{graphicx,times}             
\usepackage{natbib}
\usepackage{epsfig}
\usepackage{geometry}

\begin{document}

   \title{Analysis of sudden variations in photospheric magnetic fields during a large 
flare and their influences in the solar atmosphere}

   \volnopage{Vol.0 (200x) No.0, 000--000}      
   \setcounter{page}{1}          

   \author{Brajesh Kumar
      \inst{1}
   \and A. Raja Bayanna
      \inst{1}
   \and P. Venkatakrishnan
      \inst{1,2}
   \and Shibu K. Mathew
      \inst{1}
   }

   \institute{Udaipur Solar Observatory, Physical Research Laboratory, Dewali, Badi Road, 
              Udaipur 313 004, Rajasthan, India; {\it brajesh@prl.res.in}\\
        \and
             Indian Institute of Astrophysics, II Block, Koramangala, 
             Bangalore 560 034, India \\
   }

  \date{Received~~2009 month day; accepted~~2009~~month day}

\abstract{ The solar active region NOAA 11719 produced a large two-ribbon flare 
on 11 April 2013. We have investigated the sudden variations in the photospheric 
magnetic fields in this active region during the flare employing the magnetograms obtained 
in the spectral line Fe~I~6173~\AA~by the Helioseismic and Magnetic Imager (HMI) onboard 
{\em Solar Dynamics Observatory} ({\em SDO}). The analysis of the line-of-sight 
magnetograms from HMI show sudden and persistent magnetic field changes 
at different locations of the active region before the onset of the flare and during the 
flare. The vector magnetic field observations available from HMI also show coincident variations 
in the total magnetic field strength and its inclination angle at these locations. Using the 
simultaneous Dopplergrams obtained from HMI, we observe perturbations in the photospheric 
Doppler signals following the sudden changes in the magnetic fields in the aforementioned 
locations. The power spectrum analysis of these velocity signals show enhanced acoustic power in 
these affected locations during the flare as compared to the pre-flare 
condition. Accompanying these observations, we have also used the near-simultaneous chromospheric 
observations obtained in the spectral line H$\alpha$~6562.8~\AA~by the Global Oscillation 
Network Group (GONG) to study the evolution of flare ribbons and intensity oscillations 
in the active region. The H$\alpha$ intensity oscillations also show 
enhanced oscillatory power during the flare in the aforementioned locations. These results indicate 
that the transient Lorentz force associated with the sudden changes in the magnetic fields 
could drive the localized photospheric and chromospheric oscillations, like the flare-induced 
oscillations in the solar atmosphere.
}

   \authorrunning{Brajesh Kumar et al.} 
   \titlerunning{Sudden variations in magnetic fields during flare}  

   \maketitle

%
%
\section{Introduction}           
\label{sect:intro}

Solar flares are one of the most catastrophic events taking place in the solar atmosphere. During 
the flares, the magnetic energy stored in the corona is explosively released on short time-scales 
(tens of minutes) in the form of thermal radiations in excess of $10^{32}$ erg and also produces 
energetic particles moving with very high speeds. The magnetic field configuration in the 
corona rapidly changes during the flares and the signatures of these appear in the form of the 
evolution of the photospheric magnetic fields. The changes in the photospheric magnetic fields 
during flares take place, both on short time-scales of few minutes during the impulsive phase 
of the flare and longer time-scales of hours covering the phases before and after the flare. 
The rapid short-term magnetic field changes (magnetic transients) during flares was first of 
all reported by \cite{patt81}. However, it was interpreted later on \citep{patt84,harvey86} 
that the reported magnetic transients could be attributed to 
the transient emission of the spectral line being used for measuring the magnetic fields. 
This interpretation was further established by \cite{qiu03} through simulations 
for an observed transient polarity reversal in photospheric magnetic field measurements 
obtained by Michelson and Doppler Imager (MDI; \citealp{scherrer95}) 
onboard {\em Solar and Heliospheric Observatory} ({\em SOHO}; \citealp{domingo95}) 
spacecraft. On the other hand, \cite{wang92} and \cite{wang94,wang02}
reported rapid and permanent magnetic field changes in flaring active regions using the 
magnetic field observations at Big Bear Solar Observatory (BBSO). These 
observations were also confirmed by \cite{koso01} by analyzing the 
line-of-sight magnetograms obtained from the MDI instrument for an X-class flare on 
14 July 2001. Later on, \cite{sudol05} and \cite{petrie10} extensively analyzed the changes in 
line-of-sight magnetic fields accompanying several X- and M-class flares. They showed 
abrupt, significant, and permanent changes in the longitudinal magnetic fields in the 
flaring active regions. These changes were typically observed to take place in less than 
10 minutes with median magnitudes of 100~G. This sudden re-configuration 
of the magnetic fields during the flare produce Lorentz-force-transients in the solar 
atmosphere and is termed as ``magnetic-jerk''. \cite{hudson08} and \cite{fisher12} 
estimated that the Lorentz force 
(of the size $\sim$$10^{22}$~dynes) associated with the magnetic jerks could be 
responsible of driving localized seismic waves in the solar photosphere. \cite{kumar11} 
found good correspondence between the enhanced localized 
photospheric velocity oscillations and the sites of magnetic jerks observed in the 
solar active region NOAA 10930 during an X3.4-class flare on 13 December 2006. 
Since the magnetic field lines tied to the solar photosphere extends above in the 
corona, hence it would be important to study the effect of magnetic jerks simultaneously 
at the solar surface as well as in the higher layers of the solar atmosphere.

In this work, we present a detailed analysis of magnetic and velocity field changes 
associated with a large two-ribbon flare (of class M6.5) in the solar active region 
NOAA~11719 on 11 April 2013, using the high-resolution and high-quality co-temporal 
full-disk photospheric magnetograms and Dopplergrams obtained from the 
Helioseismic and Magnetic Imager (HMI; \citealp{schou12}) instrument 
onboard {\em Solar Dynamics Observatory} ({\em SDO}; \citealp{pes12}) spacecraft. 
Accompanying these observations, we have also used near-simultaneous full-disk 
chromospheric H$\alpha$ observations of this flare event obtained from the 
Global Oscillation Network Group (GONG; \citealp{harvey95,harvey96}) instrument 
in order to study the morphological and spatial 
evolution of the flare-ribbons as well as the H$\alpha$ intensity oscillations 
during the flare. The motivation of this work is to further investigate the 
influence of magnetic jerks on the localized photospheric velocity oscillations in the 
active region during this flare. 
We also aim to investigate the possible variations in the chromospheric H$\alpha$ 
intensity oscillations corresponding to the photospheric locations of the magnetic 
jerks in the active region during the flare. This would be useful to understand 
the physical mechanism inter-linking the different layers of the solar atmosphere.

In the following Sections, we describe the observational data, our approach of 
the data analysis, the results, and the conclusions with a discussion.

\section{The observational data}
\label{sect:Obs}

The solar active region NOAA~11719 appeared on the east limb of the Sun on 5 April 2013 in the northern 
hemisphere and slowly it evolved into a very complex magnetic region (of class $\beta\gamma\delta$). 
It produced several C- and M-class flares during its passage on the solar disk. On 11 April 2013, a 
large two-ribbon flare (of class M6.5) occurred around 06:55~UT in this complex active region when it was 
located at the heliographic coordinates 
N10E08. The flare produced a high speed Earth-directed Coronal Mass Ejection (CME), type II radio bursts and 
protons of energy more than 10 MeV. Observations related to this interesting flare are available from 
various space- and ground-based observatories in different wavelengths and energies, 
viz., {\em SDO}, {\em Hinode} \citep{kosugi07}, 
{\em Solar TErrestrial RElations Observatory} ({\em STEREO}; \citealp{kaiser08}), 
{\em Geostationary Operational Environmental Satellite-15} ({\em GOES-15}), 
Nobeyama Radio Polarimeters (NoRP; \citealp{naka85}, and references therein), and GONG. 
In our analysis related to this flare, we employ the 
photospheric velocity and magnetic field observations obtained by the Helioseismic and Magnetic 
Imager (HMI) instrument onboard {\em SDO}, 
chromospheric H$\alpha$ observations obtained from the GONG++ telescope \citep{harvey11}, 
and the soft X-ray observations 
obtained from the {\em GOES-15} satellite during the flare event.

\subsection{{\em SDO}/HMI Observations}

The HMI instrument onboard {\em SDO} spacecraft provides co-temporal, high quality full-disk 
photospheric Dopplergrams and 
line-of-sight magentograms, taken in Fe~{\sc i}~6173~\AA~line at the cadence of 45~s with a spatial 
sampling rate of $\sim$0.5~arcsec per pixel. HMI derives the line-of-sight velocity and magnetic 
fields using the Stokes $I$ and $V$ measurements obtained by imaging spectro-polarimetry. 
The HMI instrument team also provides photospheric vector magnetograms at a cadence of 12~minute 
using spectropolarimetric measurements of Stokes $I$, $Q$, $U$, and $V$. 
In this study, we have used the tracked 
grid ($\sim$290$\times$193~arcsec$^2$) of magnetic maps and Doppler 
images of the active region NOAA~11719 for the period from 01:00 UT to 10:00 UT 
on 11 April 2013 to study the flare associated changes in the photospheric magnetic and velocity 
fields in the active region. The HMI team produces the tracked data of the active region after 
remapping the region of interest onto heliographic coordinates. 
In addition to these images, we have also used the continuum images of the active region as observed with 
HMI to study its morphology during the flare. In the Figure~1, we show the mean images of the 
active region NOAA~11719 in the continuum, the line-of-sight magnetic fields, and the total Doppler velocity 
constructed over the time period from 07:00 UT to 08:50 UT on 11 April 2013. It is clearly seen in 
these images that this active region had a very complex morphology, and hence this is the 
reason that it produced several flares during its journey through the solar disk.

\subsection{GONG++ H$\alpha$ Observations}

The network of GONG instruments was upgraded to GONG+ \citep{harvey98} by replacing the 
existing CCD camera of 256$\times$256 pixels with a new CCD camera having 1024$\times$1024 pixels for better 
full-disk spatial resolution in the Dopplergrams, the line-of-sight magnetograms, and the continuum images 
acquired by this telescope. Later on in mid-2010, GONG+ was upgraded to GONG++ with the addition of 
H$\alpha$ image acquisition system to provide a nearly continuous solar activity patrol for use in space 
weather applications (Harvey et al. 2011). Since then, GONG++ network has obtained very useful 
full-disk chromospheric observations in H$\alpha$~6562.8~\AA~line at the cadence of one minute and spatial 
sampling of $\sim$1.0~arcsec per pixel. The GONG++ instrument nicely covered the full 
event of this large two-ribbon flare in H$\alpha$ along with its other 
regular observations. In the Figure~2, we show the evolution of the flare-ribbons in H$\alpha$ 
using the GONG++ observations (negative images). Here, we notice that the flare-ribbons covered 
the umbra of the sunspot 
as the flare progressed and also that it was a long duration flare event. We have used these 
H$\alpha$ observations simultaneously with the aforementioned HMI observations to study the 
co-temporal photospheric and chromospheric changes associated with this flare.

\subsection{{\em GOES-15} Soft X-ray Observations}

In order to understand the evolution of flare in the solar environment, we also require
the information related to time evolution of high energy radiations from the Sun during the flare.
In hard X-ray observations, the {\em Reuven Ramaty High Energy Solar Spectroscopic Imager} 
({\em RHESSI}; \citealp{lin02}) satellite missed to completely cover the impulsive phase of 
the flare. However, the {\em GOES-15} satellite observations in soft X-ray emissions during the flare 
are available for nearly the full event. Hence, we have used the disk-integrated soft X-ray data 
in the 1--8~\AA~band from 
{\em GOES-15} satellite for having temporal information related to the emission of high-energy radiations 
from the Sun during the flare event. In the Figure~3, we show the temporal evolution of the light-curves 
obtained from the {\em GOES-15} in 1--8~\AA~band and the H$\alpha$ (6563~\AA~line) flare-ribbons 
obtained from the GONG++ instrument. 
Here, we observe that both the high-energy light-curves peak around 07:15~UT during the flare 
event. It is also noticed that {\em GOES-15} observations have a gap of nearly 20 minutes during the 
declining phase of the flare, which would hardly affect our study.

\section{Analysis and Results}
\label{sect:results}

The chief motivation of this work is to investigate any sudden variations in the line-of-sight 
photospheric magnetic fields in the localized regions of the active 
region during the flare and their influence on 
the localized velocity oscillations on the solar surface. We are also interested to examine the 
effect of these magnetic jerks in the solar atmosphere. For this purpose, we have employed the 
co-temporal high resolution photospheric magnetic and velocity field observations from HMI instrument 
onboard {\em SDO} spacecraft and nearly simultaneous chromospheric observations in H$\alpha$ from 
GONG++ network. The detail of the data reduction and analysis are presented as follows:

\subsection{Analysis of Magnetic Field Changes in the Active Region Using HMI Data}

We have analyzed the sequence of tracked grid of photospheric line-of-sight magnetic 
images from HMI for the period from 06:00~UT to 
10:00~UT on 11 April 2013 obtained at the cadence of 45~s, in order to search for the 
sudden changes in the line-of-sight magnetic fields ($B_{los}$) in 
the active region during the flare. The field-of-view of the grid of tracked images is 
$\sim$290$\times$193~arcsec$^2$ and it mostly covers the active region. It is to be noted 
that the flare started around 06:55~UT and peaked 
around 07:15~UT on 11 April 2013 as seen in the light-curves of soft X-ray observations from 
{\em GOES-15} and H$\alpha$ observations from GONG++ (c.f., Figure~3). Hence, the analysis of 
magnetic fields during the period from 06:00~UT to 10:00~UT would be helpful in the study 
of magnetic jerks, if any, in the active region during the flare as the aforementioned 
time period comprises of the observations before, and spanning the flare. We have scanned 
all the pixels in the field-of-view of the grid of tracked line-of-sight magnetic images 
for the detection of sudden changes in $B_{los}$ in the active region. 
The following selection criteria have been applied to these magnetic images for identifying 
such locations in the active region:

(i) Considering that the noise level in the measurement of $B_{los}$ of HMI instrument 
is less than 10~G per pixel, we have considered only those pixels whose 
average $B_{los}$ over a grid of 3$\times$3 adjoining pixels is more than $\pm$~50~G in 
the given time series.

(ii) The extensive analysis of abrupt changes in longitudinal magnetic fields 
during several major flares done by \cite{sudol05} shows that the typical value of 
the abrupt changes in $B_{los}$ is about 90~G observed during those flares. Following this, 
we have considered only those pixels where the difference between 
the maximum and minimum of the average $B_{los}$ over a grid of 3$\times$3 adjoining pixels in the 
time series is more than 100~G, separately, for the positive and negative magnetic field regions.

(iii) In order to eliminate the long-term slow variations and trends in the $B_{los}$ in the sorted 
pixels with the aforementioned criteria, their time series of average $B_{los}$ over a grid of 
3$\times$3 adjoining pixels is subjected to a $\tilde\chi^2$ test by applying a linear fit to the 
aforementioned time series. It is seen that the time series with relatively high $\tilde\chi^2$ 
value shows fast variations in $B_{los}$. This criterion eliminates the 
pixels showing slow field evolution with time, and thus we remain with only those time series 
which have `step-like changes' in $B_{los}$ spanning the flare event.

(iv) \cite{sudol05} suggest that the abrupt changes in $B_{los}$ typically occur in 
less than 10 minutes while \cite{petrie10} suggest that the median duration of 
abrupt field changes is about 15 minutes. Hence, to further distinguish between normal 
field evolution and abrupt and persistent changes in the $B_{los}$ spanning the flare event, 
we have considered only those pixels where more than 100~G change in average $B_{los}$ over 
a grid of 3$\times$3 adjoining pixels occurs in less than 20 minutes.

Using the above mentioned criteria, we found six locations in the active region 
where we observe group of pixels showing sudden and persistent changes in the $B_{los}$ 
around the impulsive phase of the flare. These 
locations are shown as `K1', `K2', `K3', `K4', `K5', and `K6' indicated by yellow circles 
in the Figure~4. From this figure, we observe that these locations are situated in the 
vicinity of the H$\alpha$ flare ribbons as well as away from the flare ribbons. In the left 
panels of Figure~5, we have plotted the average values of 
$B_{los}$ over a raster of 3$\times$3 pixels in the aforementioned locations along with the 
light-curve from {\em GOES-15}. A raster of 3$\times$3 pixels is considered in order to avoid 
the possible errors in tracking of the grid of images. The centroids of these 
rasters in the different locations are shown by blue crosses in the Figure~4. The Carrington 
heliocentric longitudes and latitudes (in degree) of the aforementioned centroids are as 
follows: K1(76.91, 12.67), K2(74.06, 9.88), K3(73.49, 12.10), K4(73.64, 14.95), 
K5(68.75, 14.05), and K6(68.21, 17.26). In the plots of $B_{los}$ in the Figure~5, 
we observe in most of the cases that the value of the longitudinal magnetic fields 
in the aforementioned locations changes abruptly, then maintains the 
new value for a period of time that is significantly longer than the duration of 
the sudden changes in $B_{los}$, and then shows unsteady behaviour. Thus, these abrupt changes 
in $B_{los}$ in our observations are 
``persistent'' but not ``permanent'' changes in longitudinal magnetic fields as reported earlier by 
\cite{sudol05} and \cite{petrie10} for other flare events. It is also to be noted that 
the sudden changes in $B_{los}$ appear 
before the onset of flare as well as during the flare event, and these changes are 
seen mostly in the weak to moderate magnetic field locations ($B_{los}$$<$500~G) in the 
active region. The amount of change in $B_{los}$ ranges between 100~G and 200~G taking 
place within the time scales of 10 minutes as evident from the plots shown in the Figure~5. 

In order to understand the cause behind such variations in $B_{los}$, we 
have studied the morphological evolution of $B_{los}$ at the locations 
`K1', `K2', `K3', `K4', `K5', and `K6' in the active region during the flare. 
In the Figure~6, we show the mosaic of the time evolution of $B_{los}$ over a grid of 
15$\times$15 pixels containing the raster of 3$\times$3 pixels at its center 
(enclosed by white box) corresponding 
to the locations K1(76.91, 12.67), K2(74.06, 9.88), 
K3(73.49, 12.10), K4(73.64, 14.95), K5(68.75, 14.05), and K6(68.21, 17.26) 
in the active region for the 
period from 06:30~UT to 08:18~UT at the time interval of twelve minutes. The time 
duration from 06:30~UT to 08:18~UT covers the pre-flare phase and the flare phase. 
In this illustration of the morphological evolution of $B_{los}$, 
we observe fast changes in the magnetic concentrations
in these affected locations. We conjecture that the sudden changes in the 
$B_{los}$ observed at the aforementioned locations could be due to the process of 
fast re-organization of the coronal magnetic fields during the flare, thereby 
producing fast changes in the directions of the photospheric magnetic 
fields.

We have also used the tracked vector magnetic field data available from HMI at the 
cadence of twelve minute for the period from 06:00~UT to 10:00~UT on 11 April 2013, 
in order to study the variations in the total magnetic field strength ($B_0$) and 
the inclination angle ($\gamma$) of the 
magnetic field lines at the locations of sudden changes seen in $B_{los}$ 
in the active region. These data are co-aligned with the main grid of $B_{los}$ images 
($\sim$290$\times$193~arcsec$^2$) for estimating $B_0$ and $\gamma$ at 
the affected locations. 
In the Figure~7, we show the plots of $B_0$ and $\gamma$ averaged over the same 
rasters of 3$\times$3 pixels at the locations K1(76.91, 12.67), K2(74.06, 9.88), 
K3(73.49, 12.10), K4(73.64, 14.95), K5(68.75, 14.05), and K6(68.21, 17.26) 
in the active region as considered for $B_{los}$ (c.f., Figure~5) for the 
period from 
06:00~UT to 10:00~UT on 11 April 2013. Here, we notice that both $B_0$ and $\gamma$ 
show sudden and persistent variations around the time of the sudden changes seen in the $B_{los}$. 
This implies that the sudden and persistent changes seen in 
the $B_{los}$ are due to the simultaneous sudden changes in $B_0$ and $\gamma$ 
at the locations `K1', `K2', `K3', `K4', `K5', and `K6' in the active region. 

These sudden changes in the magnetic fields at the aforementioned locations would 
produce Lorentz-force-transients or the ``magnetic-jerk'' in these affected 
areas.

\begin{figure*}
\centering
\includegraphics[width=0.61\textwidth, angle=90]{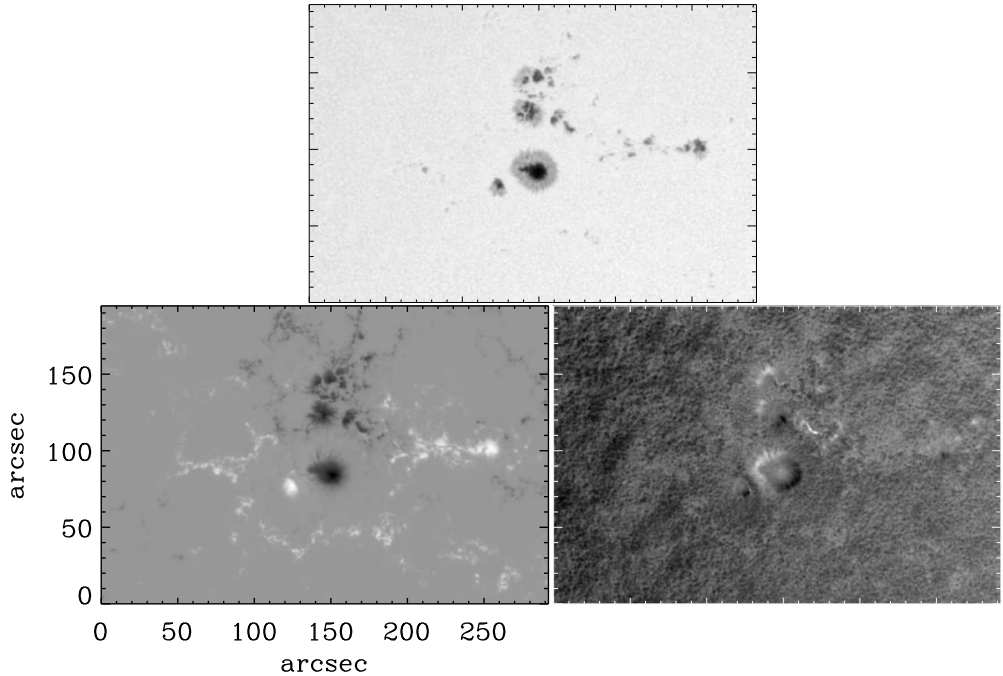}
\caption{The images of the active region NOAA~11719 in the continuum intensity (top panel), 
the line-of-sight photospheric magnetic fields (bottom left panel), and the photospheric Doppler 
velocity (bottom right panel) as observed with HMI instrument onboard 
{\em SDO} spacecraft. All the images shown here are the mean images constructed over the time interval 
07:00-08:50~UT spanning the flare on 11 April 2013.}
\end{figure*}

\begin{figure*}
\centering
\includegraphics[width=0.61\textwidth, angle=90]{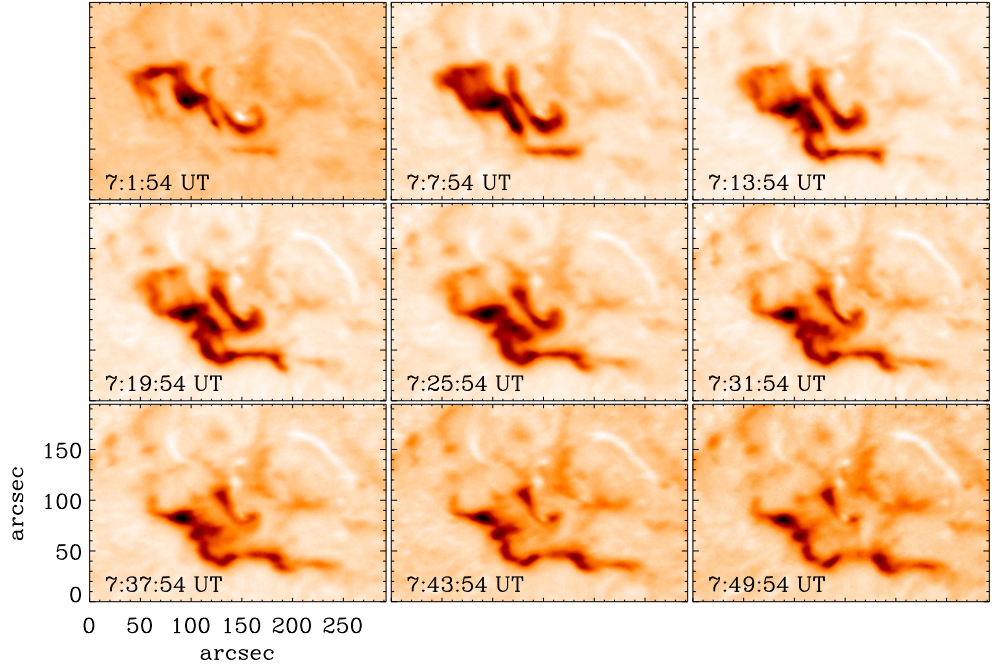}
\caption{ Mosaic of images showing the evolution of the flare-ribbons in 
H$\alpha$~6563~\AA~line during the period 07:01:54-07:49:54~UT on 11 April 2015 
using the GONG++ observations. Here, we notice that this was a two-ribbon flare and the flare-ribbons 
separated apart and covered the umbra of the sunspot as the flare progressed. The images are shown 
in negative for better appearance of the morphology of the flare-ribbons.}
\end{figure*}

\begin{figure*}
\centering
\includegraphics[width=0.52\textwidth, angle=90]{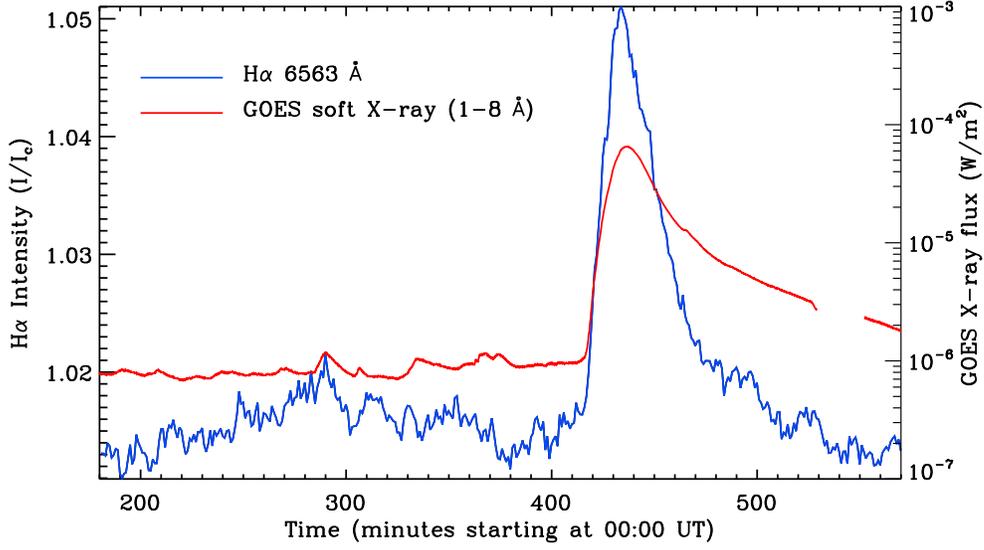}
\caption{ Plots showing the the temporal evolution of the light-curves 
obtained from the {\em GOES-15} in 1--8~\AA~band and the H$\alpha$ (6562.8~\AA~line) flare-ribbons 
obtained from the GONG++ instrument during the period 03:00-09:30~UT on 11 April 2013. 
Here, we observe that both the high-energy light-curves peak around 07:15~UT during the flare 
event. It is also noticed that {\em GOES-15} observations have a gap of nearly 20 minutes during the 
declining phase of the flare.}
\end{figure*}

\begin{figure*}
\centering
\includegraphics[width=0.52\textwidth, angle=90]{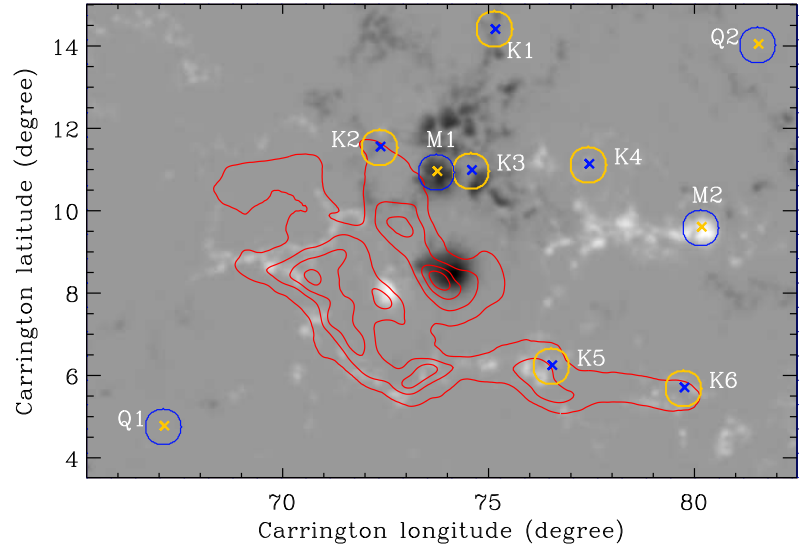}
\caption{ The background shows the mean line-of-sight photospheric 
magnetic fields for 
the period 07:00-08:50~UT on 11 April 2013 in the active region NOAA~11719, as 
measured in Fe~I~6173~\AA~line by the HMI instrument onboard {\em SDO} spacecraft. The overlaid 
contours shown in red color are the locations of the chromospheric H$\alpha$ (6562.8~\AA~line) 
flare-ribbons averaged over the period 07:00-08:50~UT on 11 April 2013 in the 
aforementioned active region 
during an M6.5 class flare, as observed with the GONG++ instrument. 
The contours are drawn at the levels 90\%, 80\%, 70\% and 60\% of the maximum brightness 
in H$\alpha$, from the center of the ribbons, respectively. 
The crosses shown in blue color within the yellow circles 
(labelled as `K1', `K2', `K3', `K4', `K5', and `K6') are the centroids of the locations 
where we observe sudden and persistent changes in the line-of-sight photospheric magnetic 
fields ($B_{los}$) in the active region during 
the flare. Here, it is evident that 
these locations are in the vicinity of the flare-ribbons and away from the 
flare-ribbons in the active region. The crosses shown in yellow color within blue circles 
(labelled as `M1', `M2') are the centroids of the locations where we observe gradual 
evolution of $B_{los}$ during the flare whereas the same labelled as 
`Q1', and `Q2' are in the quiet Sun. The field-of-view is $\sim$290$\times$193~arcsec$^2$.}
\end{figure*}

\begin{figure*}
\centering
\includegraphics[width=0.70\textwidth]{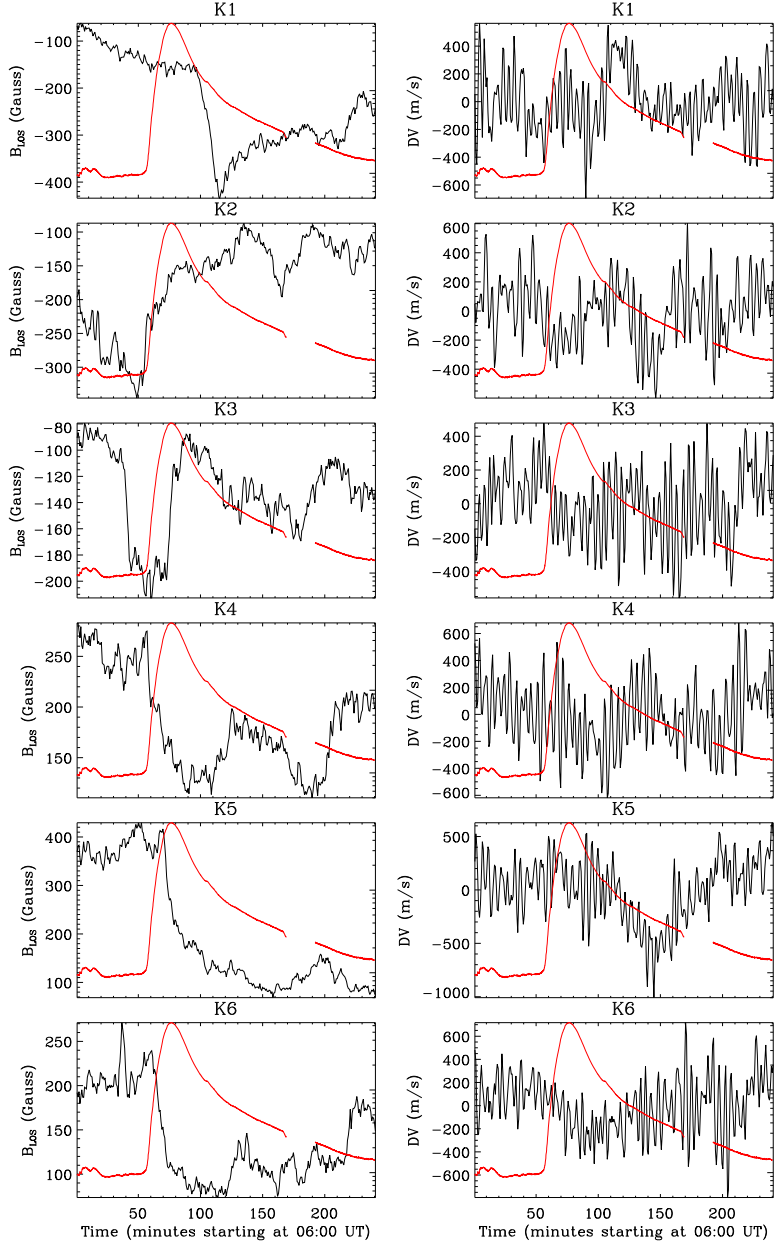}
\caption{ Plots shown in solid black lines in the left panels represent time 
evolution of the total line-of-sight photospheric magnetic field ($B_{los}$) 
averaged over a raster of nine 
pixels in the locations `K1', `K2', `K3', `K4', `K5', and `K6' in the active region  
during the period 06:00-10:00~UT on 11 April 2013 at the cadence of 45~s. Similarly, 
the plots shown in solid black lines in the right panels represent time evolution of 
the simultaneous Doppler velocity (DV) in the aforementioned locations after removing the 
large background gradient due to orbital velocity of the satellite. The $B_{los}$ 
and DV measurements are from HMI. The plots shown 
in solid red lines represent the evolution of soft X-ray flux in 1--8~\AA~energy-band 
from {\em GOES-15}. The $B_{los}$ in these locations shows sudden and persistent changes 
during the flare, which are followed by the perturbations in the DV at these locations.}
\end{figure*}

\begin{figure*}
\centering
\includegraphics[width=0.70\textwidth]{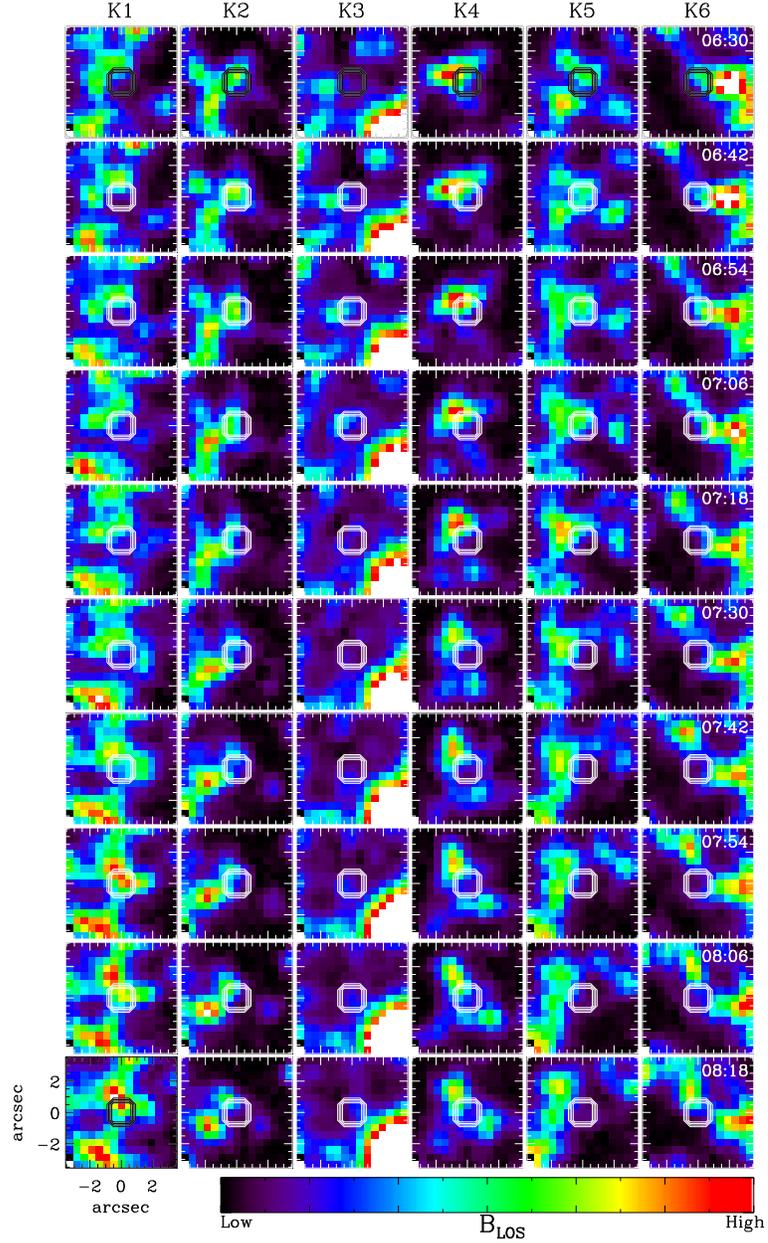}
\caption{ Mosaic of images showing the time evolution of 
$B_{los}$ over a grid of 
15$\times$15 pixels containing the raster of 3$\times$3 pixels at its center 
(enclosed by white box) corresponding to the centroids of the locations 
`K1', `K2', `K3', `K4', `K5', and `K6' in the active region for the 
period from 06:30~UT to 08:18~UT at the time interval of twelve minutes. 
Each column is representing the temporal evolution of 
$B_{los}$ for a given kernel from top to bottom. The dynamic range of $B_{los}$ 
is different for the columns (kernels), however within a column the range 
is same. For the locations `K3', and `K6', the pixels with high values of 
$B_{los}$ are made saturated in order to enhance the visibility of the 
morphological evolution of magnetic features within the white boxes. 
The FOV is $\sim$7.5$\times$7.5~arcsec$^2$.}
\end{figure*}

\begin{figure*}
\centering
\includegraphics[width=0.70\textwidth]{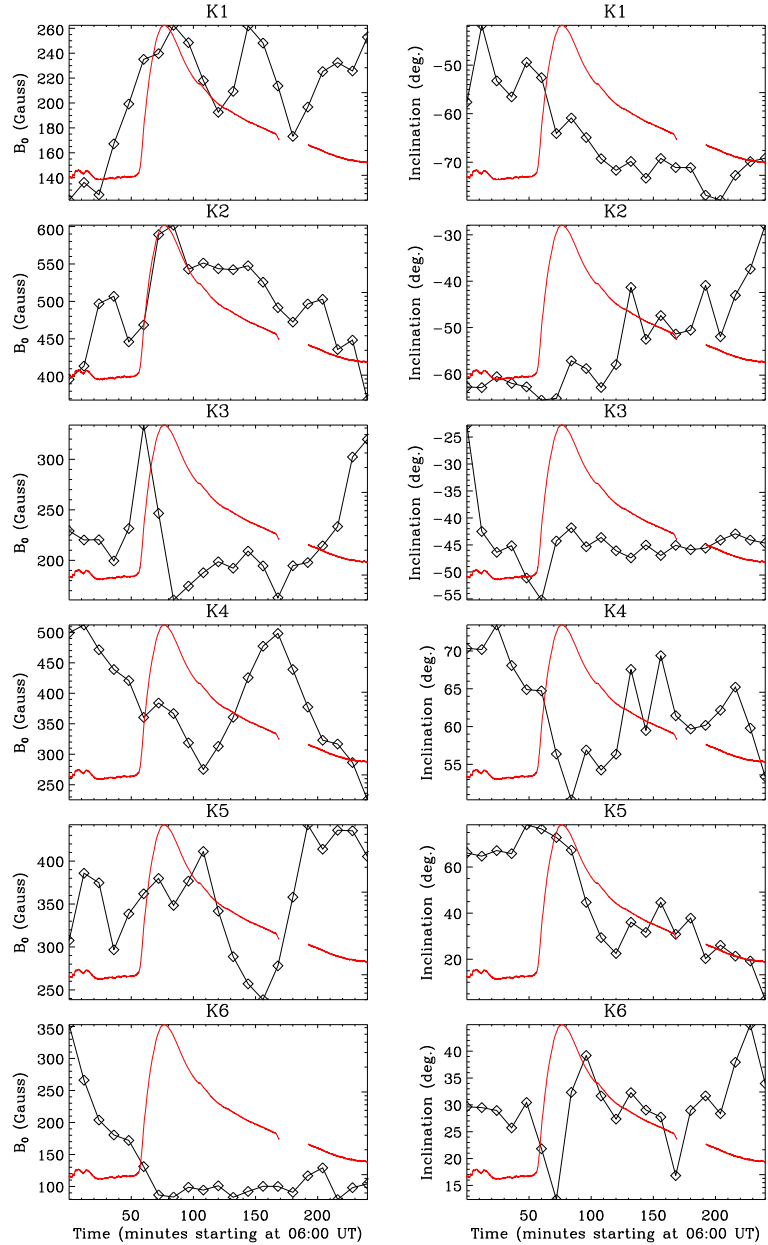}
\caption{Plots shown in solid black lines (with diamonds) in the 
left panels represent time evolution of the total 
magnetic field strength ($B_0$) averaged over a raster of nine pixels in the 
locations `K1', `K2', `K3', 
`K4', `K5', and `K6' in the active region during the period 06:00-10:00~UT 
on 11 April 2013 at the cadence of 12~minute. Similarly, the plots shown in solid black 
lines (with diamonds) in the right panels represent time evolution of 
the simultaneous inclination angle ($\gamma$) in the aforementioned locations. The $B_0$ 
and $\gamma$ measurements are from HMI. The plots shown in solid red lines in all the 
panels represent the evolution of soft X-ray flux in 1--8~\AA~energy-band as observed with the 
{\em GOES-15}. Here, we notice that the $B_0$ and the $\gamma$, both, show sudden changes 
in these locations during the flare.}
\end{figure*}

\begin{figure*}
\centering
\includegraphics[width=0.70\textwidth]{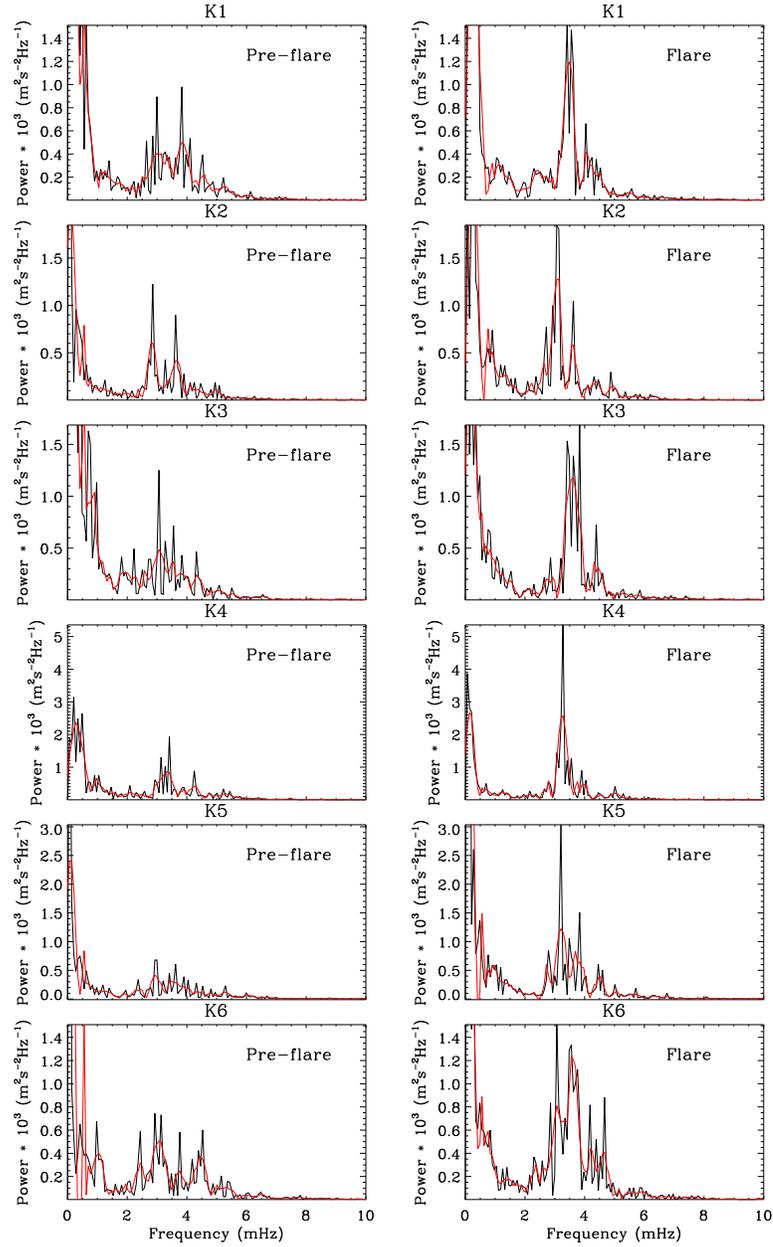}
\caption{ Plots shown in solid black lines in the left panels represent the 
average Fourier power spectrum 
of velocity oscillations estimated over a raster of nine pixels in the 
locations `K1', `K2', `K3', 
`K4', `K5', and `K6' in the active region during the period 01:00-05:00~UT 
on 11 April 2013. Similarly, the plots shown in solid black lines in 
the right panels represent 
the average Fourier power spectrum of velocity oscillations estimated over the same rasters 
of nine pixels in the 
aforementioned locations during the period 06:00-10:00~UT 
on 11 April 2013. The plots shown in solid red lines in all the panels represent a 
smoothing fit (Savitszky-Golay Fit) applied to the original power spectrum to estimate 
the power envelopes. These plots show that the acoustic power is enhanced in the 
aforementioned locations in the $2-5$~mHz frequency band during the flare.}
\end{figure*}

\begin{figure*}
\centering
\includegraphics[width=0.70\textwidth]{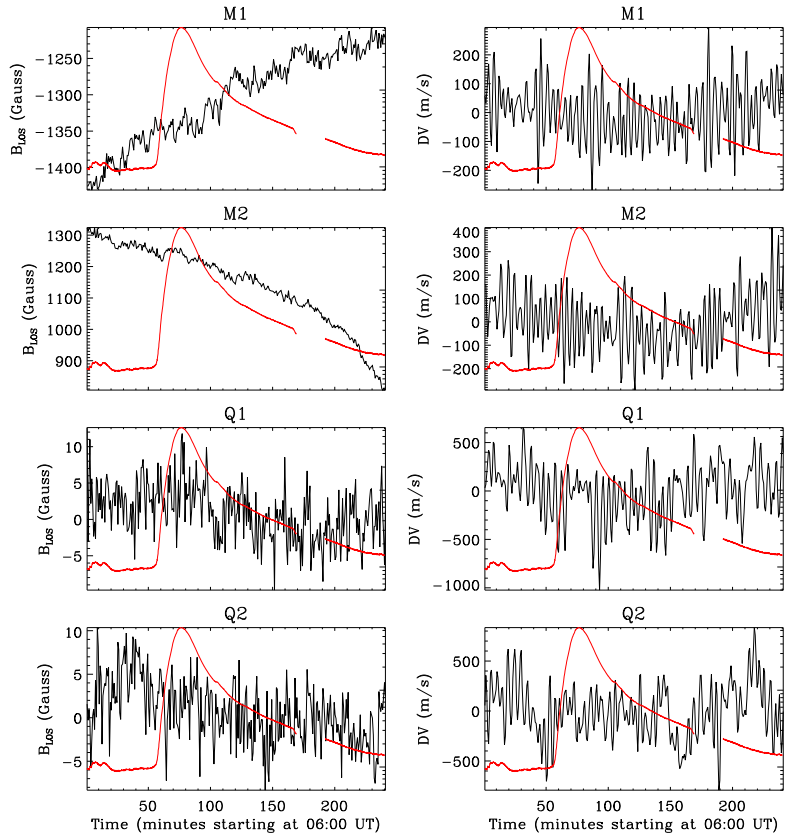}
\caption{ Same as Figure~5, but for the locations `M1', and `M2' in 
the active region and the locations `Q1', and `Q2' in the quiet Sun. Here, we notice 
that the line-of-sight photospheric magnetic 
fields ($B_{los}$) show gradual evolution in the locations `M1', and `M2' during the 
flare. We do not observe any significant changes in the photospheric 
Doppler velocity (DV) in these locations during the flare. In the quiet 
locations `Q1', and `Q2', the variations in the mean $B_{los}$ are within the 
noise level of the instrument ($\sim$10~G) while the DV is dominant and shows 
normal evolution during the flare.}
\end{figure*}

\begin{figure*}
\centering
\includegraphics[width=0.70\textwidth]{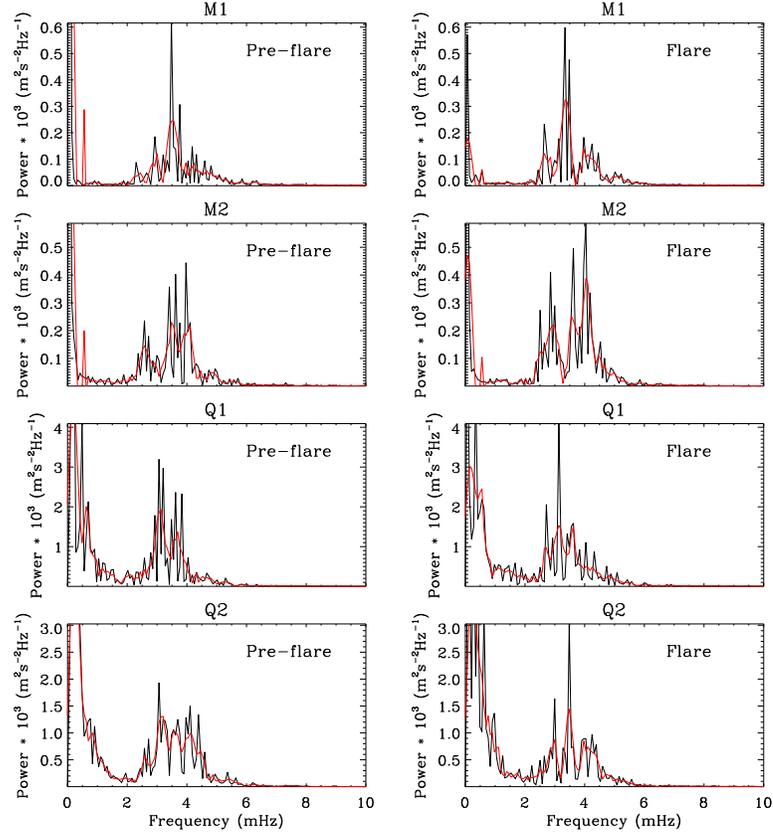}
\caption{ Same as Figure~8, but for the locations `M1', and `M2', in 
the active region and the locations `Q1', and `Q2' in the quiet Sun. Here, we notice 
that the power of velocity oscillations is enhanced in the locations `M1', and `M2' 
in the $2-5$~mHz frequency band during the flare. This enhancement in power is more for `M2' 
as compared to that for `M1' which could be attributed to the amount of reduction in the 
line-of-sight photospheric magnetic fields in the active region during the flare. 
In the case of `Q1', and `Q2', we do not observe any significant variations in the power of 
velocity oscillations during the flare, being in the quiet Sun.}
\end{figure*}

\begin{figure*}
\centering
\includegraphics[width=0.70\textwidth]{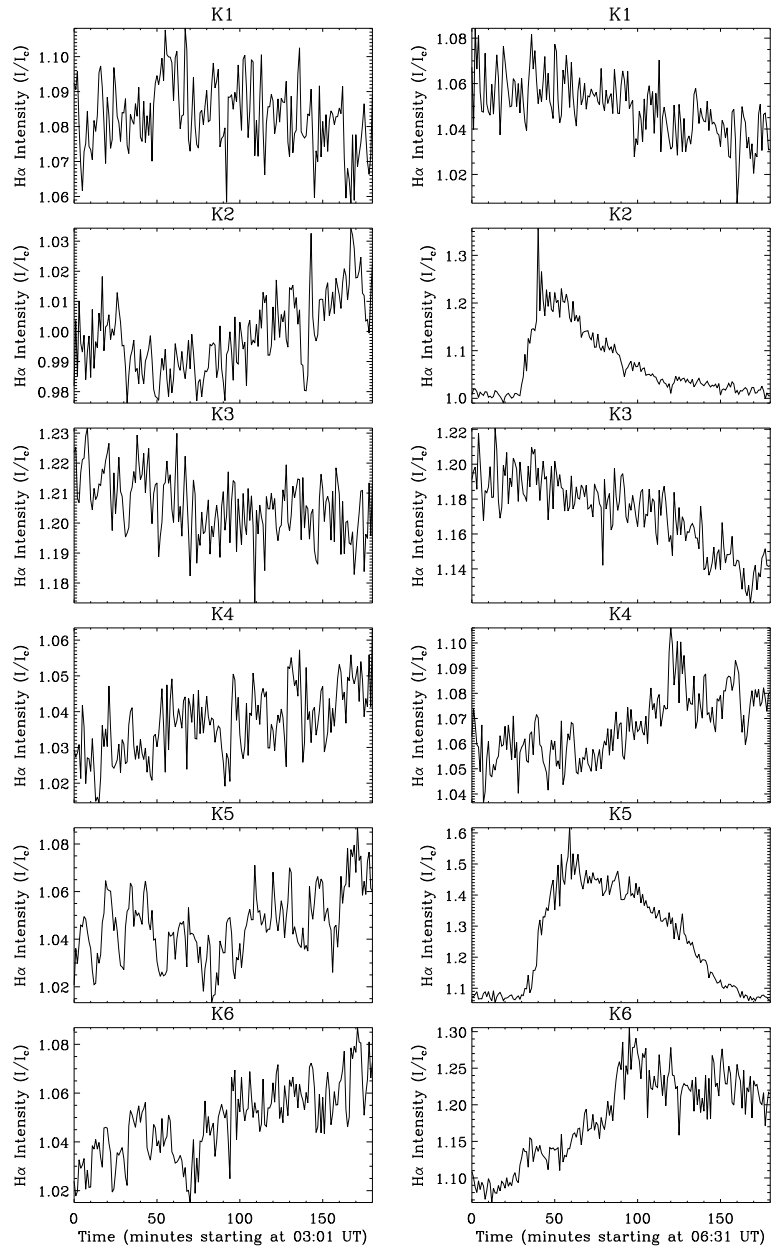}
\caption{Plots shown in solid black lines in the left panels 
represent the time evolution of 
normalized H$\alpha$ light-curves averaged over a raster of nine pixels 
within the locations `K1', `K2', `K3', 
`K4', `K5', and `K6' in the active region during the period 03:01-06:01~UT 
on 11 April 2013 (pre-flare condition). Similarly, the plots shown in solid black lines 
in the right panels represent 
the time evolution of normalized H$\alpha$ light-curves averaged over 
the same rasters of nine pixels 
within the aformentioned locations during the period 06:31-09:31~UT on 11 April 2013 
(spanning the flare). Here, the observational cadence is one minute.}
\end{figure*}

\begin{figure*}
\centering
\includegraphics[width=0.70\textwidth]{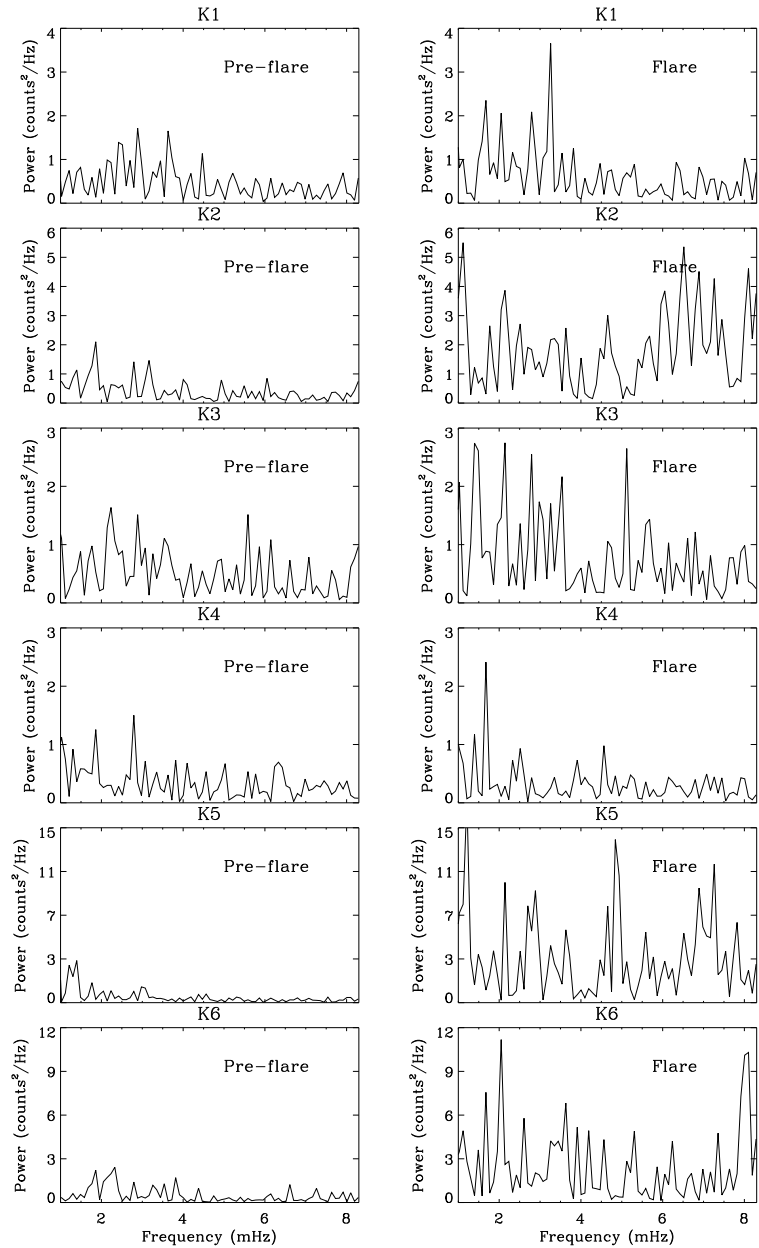}
\caption{Plots shown in solid black lines in the left panels show the 
average Fourier power spectrum 
of H$\alpha$ intensity oscillations estimated over a raster of nine pixels within the 
locations `K1', `K2', `K3', 
`K4', `K5', and `K6' in the active region during the period 03:01-06:01~UT 
on 11 April 2013 (pre-flare condition). Similarly, the plots shown in solid black lines 
in the right panels show 
the average Fourier power spectrum of H$\alpha$ intensity oscillations estimated over the 
same rasters of nine pixels within the 
aforementioned locations of the active region for the period 06:31-09:31~UT 
on 11 April 2013 (spanning the flare). It is evident that the power of H$\alpha$ 
intensity oscillations is enhanced in the aforementioned locations during the flare.}
\end{figure*}

\begin{figure*}
\centering
\includegraphics[width=0.70\textwidth]{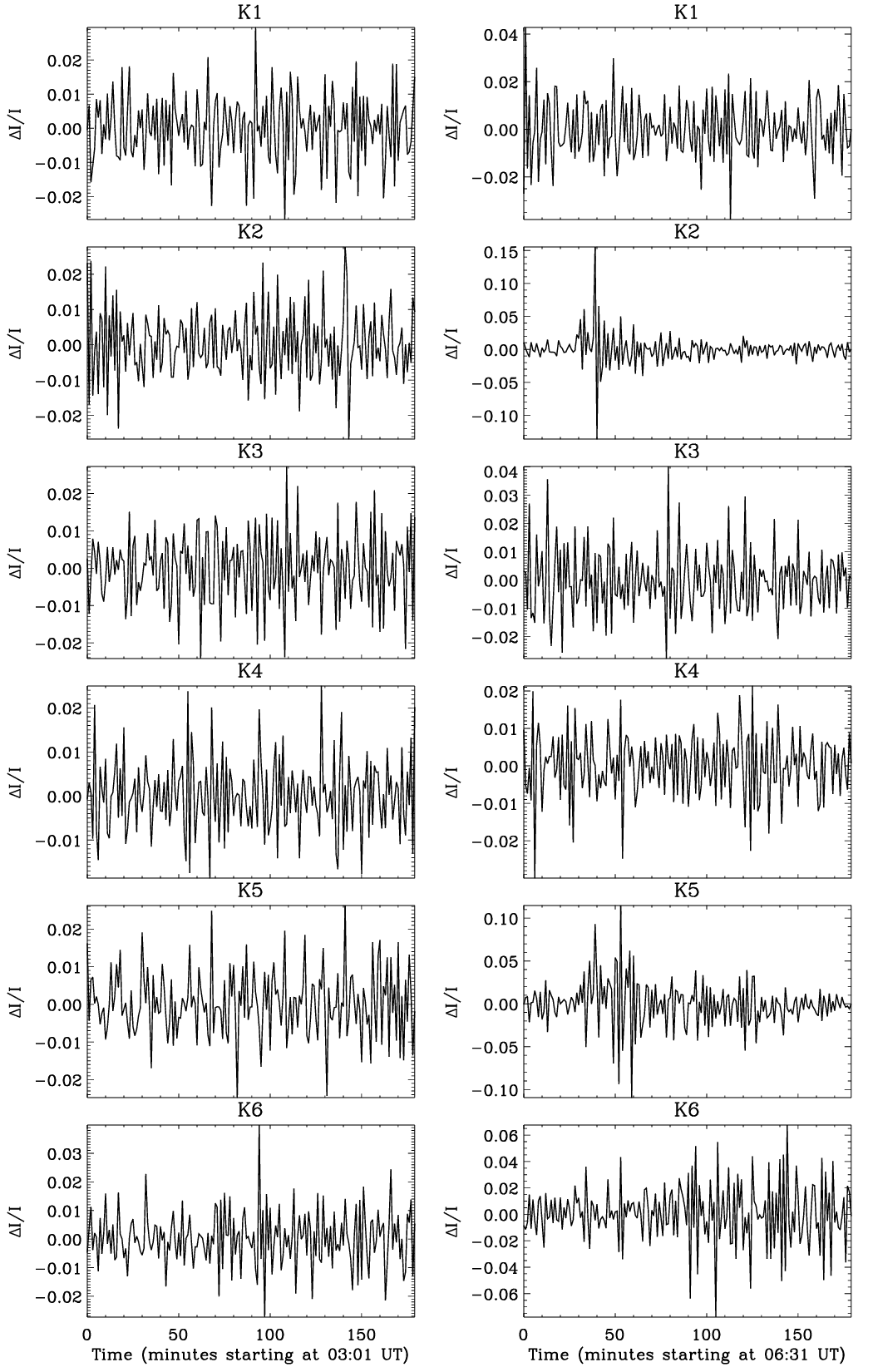}
\caption{Plots shown in solid black lines in the left panels 
represent time evolution of 
$\Delta$I/I (extracted from H$\alpha$ light-curves) averaged over a raster of nine pixels 
within the locations `K1', `K2', `K3', 
`K4', `K5', and `K6' in the active region during the period 03:01-06:01~UT 
on 11 April 2013 (pre-flare condition). Similarly, the plots shown in solid black lines 
in the right panels represent 
the time evolution of $\Delta$I/I (extracted from H$\alpha$ light-curves) averaged over 
the same rasters of nine pixels 
within the aformentioned locations during the period 06:31-09:31~UT on 11 April 2013 
(spanning the flare). The oscillatory behaviour of the quantity $\Delta$I/I 
(proxy for chromospheric velocity amplitude) is clearly 
seen in these plots.}
\end{figure*}

\begin{figure*}
\centering
\includegraphics[width=0.70\textwidth]{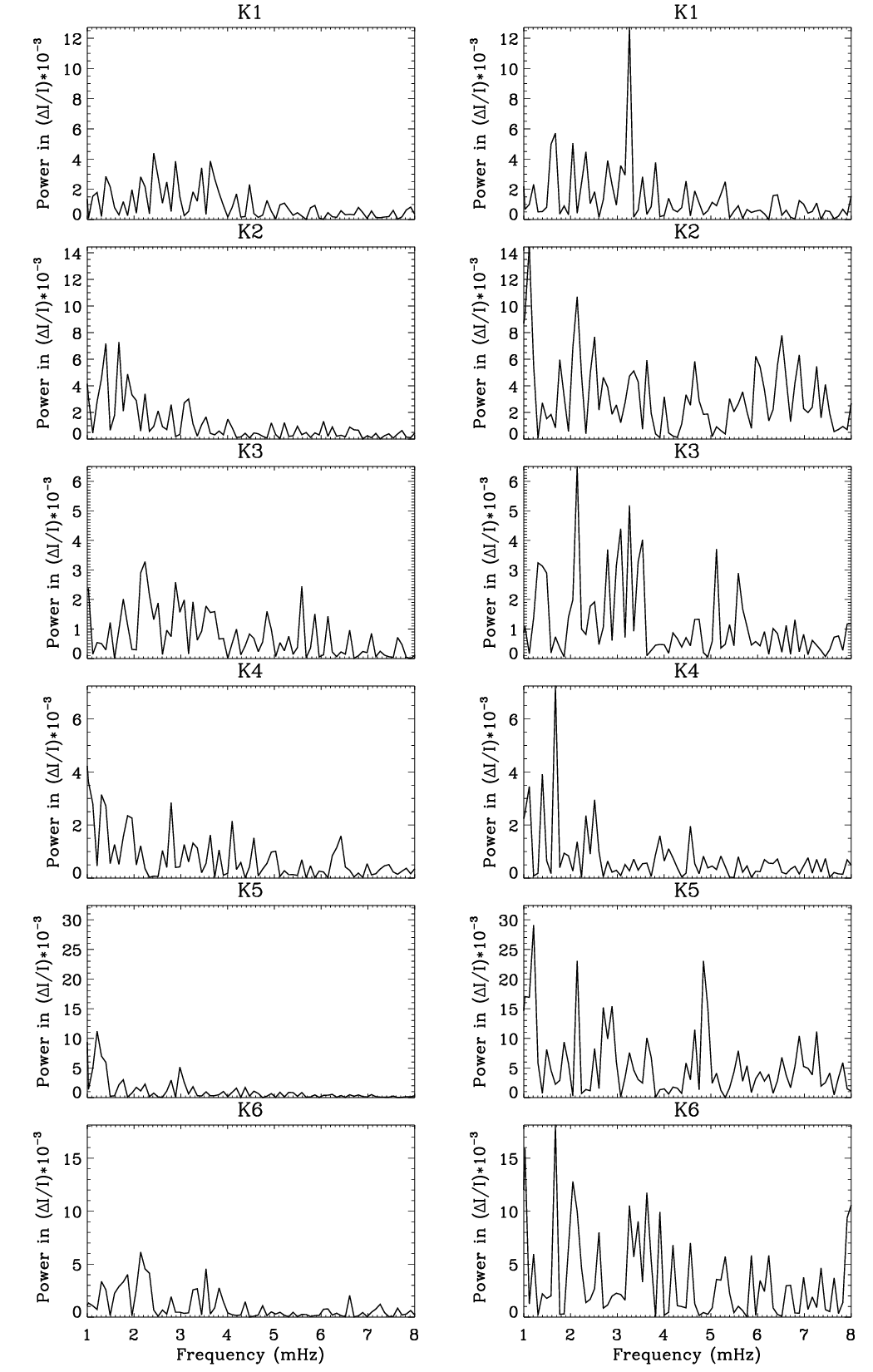}
\caption{Plots shown in solid black lines in the left panels show the 
average Fourier power spectrum of $\Delta$I/I (extracted from H$\alpha$ light-curves) 
estimated over a raster of nine pixels within the 
locations `K1', `K2', `K3', 
`K4', `K5', and `K6' in the active region during the period 03:01-06:01~UT 
on 11 April 2013 (pre-flare condition). Similarly, the plots shown in solid black lines 
in the right panels show 
the average Fourier power spectrum of $\Delta$I/I (extracted from H$\alpha$ light-curves) 
estimated over the 
same rasters of nine pixels within the 
aforementioned locations of the active region for the period 06:31-09:31~UT 
on 11 April 2013 (spanning the flare). It is evident that the power of $\Delta$I/I 
oscillations (proxy for chromospheic velocity oscillations) is enhanced in the 
aforementioned locations during the flare.}
\end{figure*}

\subsection{Analysis of Velocity Field Changes in the Active Region Using HMI Data}

The co-temporal full-disk photospheric Dopplergrams and line-of-sight magnetograms obtained 
by HMI provides the excellent opportunity to examine the changes in the velocity signals 
associated with the sites of magnetic jerks (`K1', `K2', `K3', `K4', `K5', and `K6') in 
the active region. Therefore, we have analyzed the 
sequence of tracked grid ($\sim$290$\times$193~arcsec$^2$) of Doppler velocity images from 
HMI for the period from 06:00~UT to 10:00~UT on 11 April 2013 at the cadence of 45~s. 
In our analysis, we have studied the temporal evolution of the mean velocity 
signals from the aforementioned images over the same grids of 3$\times$3 pixels at 
the centroid of the sites of magnetic jerks in the active region with the Carrington 
heliocentric longitudes and latitudes (in degree) as follows: 
K1(76.91, 12.67), K2(74.06, 9.88), K3(73.49, 12.10), K4(73.64, 14.95), 
K5(68.75, 14.05), and K6(68.21, 17.26). In the 
right panels of the Figure~5, we have plotted these mean line-of-sight velocity signals for 
the period from 06:00~UT to 10:00~UT on 11 April 2013 after removing the large background 
gradient in the observational data appearing mainly due to the orbital velocity of the satellite. 
Here, we observe significant perturbations in the photospheric Doppler signals following the epochs 
of the sudden and persistent changes seen in $B_{los}$ in the 
aforementioned locations of the active region during the flare. The signature of enhancements 
seen in the velocity signals associated with these sites of magnetic jerks motivated us to peform 
power spectrum analysis of these velocity variations during the pre-flare 
phase (01:00-05:00~UT) and the flare phase (06:00-10:00~UT) covering the magnetic jerks. 
Therefore, we applied Fourier transform to these line-of-sight velocity signals 
from each individual pixels in the grid of 3$\times$3 pixels and then an average power 
spectrum of velocity oscillations is constructed for each of the aforementioned rasters in 
the locations `K1', `K2', `K3', `K4', `K5', and `K6' in the active region for the 
pre-flare phase and the flare phase. In the Figure~8, we show in the solid black lines 
the average power spectra of 
velocity oscillations for the aforementioned locations in the active region for 
pre-flare and flare conditions. A smoothing fit 
(Savitszky-Golay Fit; \citet{press92}) is applied 
(shown in the solid red lines) to these original power spectra for estimating the 
power envelopes.

It is evident from these velocity power spectra that 
there is enhancement of power of velocity oscillations in the $2-5$~mHz band in the 
aforementioned locations during the flare as compared to the pre-flare condition. 
In this context, it is worthy mentioning that if the magnitude of the magnetic flux 
of the selected region changes, 
this would affect the amplitude of velocity oscillations in those regions leading to 
a change in the power of acoustic spectra. Thus, if there is a gross decrease in the 
magnitude of magnetic flux in the selected region during the flare, the velocity power 
spectra would show an increase of power with respect to that in the pre-flare condition 
and vice-versa. 
Hence, the method of comparing pre-flare velocity power spectrum with respect to that 
estimated during the flare may lead to erroneous interpretations. Therefore, we have checked 
the magnitude of $B_{los}$ averaged over the aforementioned grids of 3$\times$3 pixels 
in the locations 
`K1', `K2', `K3', `K4', `K5', and `K6' during the pre-flare and flare epochs considered in our 
analysis. It is seen that there is a gross increase in the magnitude of average $B_{los}$ 
in the locations `K1', and `K3' while there is a gross decrease in the magnitude of average 
$B_{los}$ in the locations `K2', `K4', `K5', and `K6' during the flare phase as compared to the 
pre-flare phase. Also, it is to be noted that `K1' and `K3' are much away from the flare-ribbons. 
However, we observe an enhancement in the power of velocity oscillations in all the 
locations `K1', `K2', `K3', `K4', `K5', and `K6'. This implies that the effect of 
``magnetic-jerk'' 
dominates the effect of the normal evolution of magnetic concentrations in the aforementioned 
locations concerning the enhancement of power of the velocity oscillations in these 
identified locations. 

In order to further verify our aforementioned observations and the interpretations, 
we have done similar analysis for some other locations in and around this active 
region for the pre-flare phase and the flare phase. For this purpose, we have 
considered two locations (`M1', and `M2') in the active region which show gradual 
evolution of $B_{los}$ during the flare, and two locations (`Q1', and `Q2') in the 
quiet Sun. The centroids of these locations are shown by yellow crosses within 
blue circles in the Figure~4. The Carrington 
heliocentric longitudes and latitudes (in degree) of the aforementioned centroids are as 
follows: M1(73.76, 10.96), M2(80.18, 9.61), Q1(67.13, 4.78), 
and Q2(81.56, 14.05). In the Figure~9, we have plotted the mean 
$B_{los}$ (left panels) and the mean line-of-sight velocity signals (right panels)  
after removing the large background gradient over a raster of 3$\times$3 pixels 
for the aforementioned kernels during the period from 06:00~UT to 10:00~UT on 
11 April 2013. In the case of `M1', we 
observe a gross decrease of $\sim$200~G in the $B_{los}$ whereas `M2' shows a 
gross decrease of $\sim$500~G in the $B_{los}$ during the flare. Here, we 
do not notice any significant perturbations in the photospheric Doppler signals 
during the flare in these locations of the active region. In the quiet 
locations `Q1', and `Q2', the 
variations in the mean $B_{los}$ are within the noise level of the instrument 
($\sim$10~G) while the mean line-of-sight velocity signals are 
dominant and show normal 
evolution during the flare. In the Figure~10, we show the average power spectra of 
velocity oscillations in these locations for pre-flare phase (01:00-05:00~UT) 
and the flare phase (06:00-10:00~UT). In the 
case of `M1', and `M2', we notice from these velocity power spectra that 
there is enhancement of power of velocity oscillations in the $2-5$~mHz band 
during the flare as compared to the pre-flare condition. We also observe that 
the enhancement in oscillatory power at the location `M2' is more as compared 
to that in the location `M1'. Here, it is worthy mentioning that during the flare 
the gross decrease in 
$B_{los}$ is more at `M2' as compared to that in `M1' and hence this has resulted 
into relatively more enhancement of oscillatory power at `M2'. In the case of 
`Q1', and `Q2', we do not observe any significant variations in the power of 
the velocity oscillations during the flare as compared to the pre-flare condition. 
These results show that the evolution of magnetic concentrations in the locations 
`M1', and `M2' is controlling the enhancement of the power of the velocity oscillations 
in these locations during the flare, while `Q1', and `Q2' remain unaffected being 
in the quiet Sun. However, as we have discussed earlier this assumption is not applicable 
at the sites of the magnetic jerks in the active region during the flare. 
We have observed enhancements in the acoustic power at the locations of magnetic 
jerks irrespective of the increase or decrease of $B_{los}$ at these locations 
during the flare with respect to the pre-flare condition. It 
is also worthy to note that the effect of magnetic jerks is more pronounced 
on the power of acoustic oscillations as compared to that due to the 
normal variations in the $B_{los}$ at the locations `M1', and `M2'.

The another important point to mention is that we do not observe any instantaneous 
large velocity signal appearing 
right at the time of the impulsive phase of the flare as has been seen in the active 
regions during several major X-class flares. Such large spiky velocity and magnetic signals 
may also appear due to distortions in the line profile in the flaring regions. In our case, 
the identified locations are in the vicinity of the flare-ribbons and also away from 
the sites of these ribbons. We observe relatively enhanced photospheric 
velocity signals following the phenomena of sudden and persistent changes seen 
in $B_{los}$ in the active region and this 
is the reason why these lead to enhancement in the gross power of the acoustic spectra 
for the flare phase covering the magnetic jerks, as compared to the pre-flare phase.

\subsection{Analysis of H$\alpha$ Intensity Oscillations in the Active Region using GONG++ Data}

We have near-simultaneous sequence of chromospheric observations consisting of full-disk 
H$\alpha$ filtergrams 
available from GONG++ along with the above magnetic and velocity field observations from HMI. 
The availability of these H$\alpha$ observations from GONG++ provides the opportunity to study 
the possible changes in the H$\alpha$ intensity oscillations driven by the aforementioned 
``magnetic-jerk'' in the active region appearing during the flare. We have reasonably 
good quality full-disk H$\alpha$ images available from GONG++ instrument for the period 
from 03:01~UT to 09:31~UT on 11 April 2013 at a cadence of one minute. The sequence of these 
every minute H$\alpha$ images are derotated and registered with respect to the image 
taken at 03:01~UT with an accuracy of 0.1 arcsec. 
A grid of the size $\sim$290$\times$193~arcsec$^2$ consisting of the active region is selected
from each of the derotated and registered sequence of the full-disk H$\alpha$ images. The 
sequence of these images are aligned with respect to the image at 03:01~UT
using Fast Fourier transform based cross-correlation algorithm with an accuracy 
of 0.1 arcsec. These registered images are re-scaled to the spatial scale of
0.5 arcsec per pixel in order to match with the pixel scale of HMI data.
The re-scaled image sequence is then co-aligned with respect to the field-of-view of 
the grid of magnetic and velocity images from HMI used in our analysis. The sequence 
of these co-aligned 
grid  of H$\alpha$ images are segregated into two parts for 
our analysis: 
the pre-flare phase (03:01-06:01~UT) and the flare phase (06:31-09:31~UT). Now, it 
would be reasonable to consider that the locations of the ``magnetic-jerk'' seen in the 
photosphere would not  
be the same in the chromosphere due to the inclination of the magnetic field lines. 
Hence, we have scanned through the pixels around the centroids of the grid of 
3$\times$3 pixels in the 
locations `K1', `K2', `K3', `K4', `K5', and `K6' as considered in the case 
of HMI magnetic and velocity images in order to find the best locations showing any 
``magnetic-jerk'' driven chromospheric oscillations in the 
active region during the flare. Following this, we found such locations in these H$\alpha$ images 
which show enhancement of power in intensity oscillations during the flare and 
are situated little away (within 2 arc-sec) from the aforementioned centroids of the 
sites of magnetic jerks in the photospheric magnetograms in the locations `K1', `K2', `K3', `K4', 
`K5', and `K6'. However, these new locations seen in H$\alpha$ images do not exactly 
match with the deviations as estimated ($\approx$~$h$$\cdot$$tan$($\gamma$)~;~$h$=difference 
in the average formation heights of the lines) from the inclination angles 
($\gamma$) available from the photospheric vector magnetograms from HMI. The most 
important reason for this mis-match could be the canopy effect in the chromosphere, where 
the magnetic field lines spread out and become more tilted in general than in the photosphere. 
However, the other probable sources of errors could be as follows: (i) the seeing effects 
in the H$\alpha$ ground-based 
observations obtained from GONG++ as compared with the space-based observations with 
HMI, (ii) the errors in the co-alignment of the images from the two different 
instruments with different spatial resolutions, and (iii) the difference in the available 
observational cadence of $B_{los}$ (45~s) and $\gamma$ (12~minute).
In the Figure~11, we show the evolution of normalized H$\alpha$ light curves averaged over a 
grid of 3$\times$3 pixels in the new locations within `K1', `K2', `K3', `K4', `K5', and `K6' 
in the active region during the 
phases of pre-flare and flare conditions. These H$\alpha$ light curves exhibit quasi-periodic 
intensity variations which are similar to the observations of \cite{wang00} for a C5.7 flare that 
ocurred in the solar active region NOAA~8673 on 23 August 1999. \cite{wang00} used 
high-resolution H$\alpha$ observations of this C-class flare obtained 
at BBSO. Resembling the velocity power spectrum, we have estimated the Fourier power 
spectrum of the H$\alpha$ intensity oscillations in each of the individual pixels in the 
new grids of 3$\times$3 pixels in the aforementioned locations and constructed average 
power spectra for the pre-flare and flare phases (c.f., Figure~12). It is evident 
from these power spectra that there is enhancement of power in chromospheric 
intensity oscillations in these locations during the flare as compared to the 
pre-flare condition.

\section{Discussion and Conclusions}
\label{sect:discussion}

We have investigated the photospheric magnetic and velocity field changes in the active region 
NOAA 11719 during a large two-ribbon flare (of class M6.5) that occurred on 11 April 2013, using 
the high-quality observations obtained from the HMI instrument onboard {\em SDO} spacecraft. 
Accompanying these photospheric observations, we have also analyzed the near-simultaneous 
chromospheric H$\alpha$ observations from the GONG++ instrument in order to understand any 
possible inter-linked physical process taking place between the different layers in the 
solar atmosphere. The chief findings of our investigations and the interpretations of 
our results are as follows:  

i. The analysis of the line-of-sight magnetic fields ($B_{los}$) from HMI shows sudden 
and persistent magnetic field changes at different 
locations of the active region during the flare. These sites (`K1', `K2', `K3', `K4', `K5', 
and `K6') are located 
in the vicinity of the H$\alpha$ flare ribbons, as well as away from the flare ribbons in the active 
region. We also observe that these abrupt changes in ($B_{los}$) 
appear before the onset of 
flare, as well as during the flare and are seen mostly in the weak to moderate magnetic field 
concentrations in the 
active region. \cite{burt15} analyzed several X-class flares and have reported 
abrupt magnetic field changes which take place before or around the 
start time of the flare. They have also shown that these field changes and the 
foot-points of the hard X-ray emission during the flare are not always co-spatial. 
Our results show that the amount of change in $B_{los}$ is between 100~G and 200~G 
taking place within the 
time scales of 10~minutes, which are similar to the findings of \cite{sudol05} and 
\cite{petrie10}. However, the abrupt changes in $B_{los}$ in our observations are 
``persistent'' but not ``permanent'' changes in longitudinal magnetic fields as have been found by 
\cite{sudol05} and \cite{petrie10} for other flare events. From our analysis related 
to the morphological evolution of $B_{los}$ at 
these locations, we conjecture that the sudden changes in the 
$B_{los}$ observed at these locations could be due to the process of 
fast re-organization of the cornal magnetic fields during the flare resulting into fast 
changes in the directions of photospheric vector fields.
We have also analyzed the vector magnetic field data available from HMI which show 
that the sudden changes seen in the $B_{los}$ are accompanied with the coincident sudden 
changes in $B_0$ and $\gamma$ 
at these locations in the active region. There have been always a concern regarding the 
distortions in the line-profile affecting the observations of magnetic and velocity fields 
in the active region during the impulsive phase of the large flares \citep{qiu03}. However, 
this speculation could be applicable for the locations situated within the flare-ribbons or 
the hard X-ray foot-points. In such cases, large short-lived changes appear in the magnetic 
and velocity fields during the flare. In our case, some of the affected locations (`K1', 
`K3', and `K4') are away from the 
H$\alpha$ flare ribbons and hence these are free from the possible effect of distortions in the 
line profile. On the other hand, for the locations in the vicinity of H$\alpha$ flare 
ribbons (`K2', `K5', and `K6'), we do not observe any large spiky signals appearing for 
short-times in the $B_{los}$ and Doppler velocity signals at these locations. It is also 
important to note that the profiles of $B_{los}$ and Doppler velocity for all 
the locations `K1', `K2', `K3', `K4', `K5', and `K6' appear similar in nature. Hence, 
we suggest that the sudden changes appearing in $B_{los}$ and the subsequent perturbations 
in the Doppler velocity signals in the aforementioned locations are real changes happening 
in the active region during the flare.

ii. It is believed that the Lorentz-force-transients associated with the ``magnetic-jerk'' can 
drive localized seismic waves in the solar photosphere \citep{hudson08,fisher12}. 
Hence, we have 
analyzed the HMI Dopplergrams for the photospheric velocity signals in the locations 
`K1', `K2', `K3', `K4', `K5', and `K6' for the epochs, before and spanning the flare. For this 
purpose, we have estimated Fourier power spectrum of these velocity oscillations in the 
aforementioned locations. It is observed that the power of velocity oscillations is enhanced 
in the locations of magnetic jerks spanning the flare as compared to the pre-flare condition. 
This enhancement is found for all the sites of magnetic jerks in the active region 
during the flare. There are other physical processes responsible for driving oscillations 
in the solar atmosphere during the flare, mainly the chromospheric shocks propagating through 
the photosphere into the solar interior \citep{fisher85} or the high-energy particle
beam impinging on the solar photosphere (\citealp{venkat08}, and references therein). However, 
this is applicable for the situation when the affected areas are within the hard X-ray 
foot-points or H$\alpha$ flare kernels.
In our case, the affected locations are situated in the vicinity 
of the flare ribbons and also much away from the flare ribbons in the active region, hence 
this indicates that magnetic jerks have certainly driven these 
localized photospheric velocity oscillations. A general estimate is that the transient Lorentz 
forces of the size $\sim$$10^{22}$~dynes \citep{hudson08,fisher12} 
associated with these magnetic jerks could be responsible for driving 
localized seismic waves in the solar photosphere. Following \cite{fisher12}, we estimate 
the approximate transient Lorentz force ($B_{los}$$\cdot$$\delta$$B_{los}$$\cdot$A/4$\pi$) 
corresponding to the 
sites of the magnetic jerks in the active region during this flare event. 
In our case, the mean value of the line-of-sight magnetic field changes ($\delta$$B_{los}$) 
is $\sim$150~G and the mean value of $B_{los}$ at these locations is $\sim$300~G over an 
area (A) of $\sim$$10^{16}~cm^{2}$ as obtained from HMI magnetograms in the 
locations of the magnetic jerks. Hence, the approximate estimate of the transient 
Lorentz force at these sites of magnetic jerks turns out to 
be $\sim$$10^{19}$~dynes. This is a lower available budget of the transient Lorentz force as compared 
to the required size of the force for driving the localized seismic waves. It is to be 
noted that the expression of \cite{fisher12} for estimating the transient Lorentz force is 
applicable for strong magnetic field regions ($\ge$~1~kG optimally at $\tau$=1 opacity layer at the 
wavelength 5000~\AA) and also approximated under 
some theoretical assumptions (For a detailed discussion, please refer to \citealp{petrie14}). 
In our case, the sites of magnetic jerks are located in 
weak to moderate field regions ($\le$~500~G) and also our estimate suffers from the 
non-availability of all the vector components simultaneously. Hence, the above constraints could be the 
reason for the deficit budget of transient force appearing in our estimation, by using the 
expression of \cite{fisher12} under approximations. However, we do 
observe enhancements in the power of velocity oscillations in the locations of the 
magnetic jerks in the active region during the flare. In our observations, the average change in 
the amplitude of localized velocity oscillations ($\Delta$$v$) associated with the magnetic jerks 
is $\sim$200~m/s. Hence, the estimate of the transient force required to drive these 
acosutic oscillations could be $\approx$~$\rho$$\cdot$($\Delta$$v^{2}$)$\cdot$A, where 
`$\rho$' is the mean density of the solar photosphere ($\sim$2$\times$$10^{-7}$$g$$cm^{-3}$). 
This yields a required budget of the driving force of $\sim$8$\times$$10^{17}$~dynes against the 
available budget of the transient force of $\sim$$10^{19}$~dynes as estimated from the 
magnetic jerks. Thus, from the aforementioned calculation it appears that the magnetic 
jerks at the locations `K1', `K2', `K3', `K4', `K5', and `K6' are capable of 
driving the localized photospheric acoustic oscillations.

iii. Using the H$\alpha$ chromospheric observations of this flare from GONG++ instrument, we have 
analyzed the changes in the H$\alpha$ intensity oscillations in the corresponding locations of 
the magnetic-jerks in the active region during the flare. The motivation of this analysis is 
to study the possible inter-linking physical processes between the different layers of the 
solar atmosphere. For this purpose, we have applied Fourier transform to the H$\alpha$ intensity 
oscillations in these affected locations for the epochs, before and spanning the flare. It is 
observed that there is enhancement of the power of H$\alpha$ intensity oscillations in these locations 
during the flare. The power enhancement is seen at different frequencies for the kernels 
`K1', `K2', `K3', `K4', `K5', and `K6'. These findings are similar to 
the results of \cite{wang00} in the case of flare driven H$\alpha$ intensity oscillations. 
In our case, some of the locations are in the vicinity of the flare-loops and hence 
the flare would have mainly contributed for the enhancement of the chromospheric intensity 
oscillations in these locations. However, for the locations which are away from the flare 
ribbons, the contribution of the magnetic jerks in driving these oscillations cannot be 
disregarded. We investigate this further by extracting the approximate estimates of 
chromospheric velocity amplitudes from $\Delta$I/I of the H$\alpha$ light-curves assuming 
that these intensity oscillations are wholly caused by the Doppler shift in the H$\alpha$ 
line due to the motion of chromospheric plasma. Under this assumption, the quantity 
$\Delta$I/I could be the proxy for the chromospheric velocity amplitude. We plot this 
quantity for all the locations `K1', `K2', `K3', `K4', `K5', and `K6' in the Figure~13. 
It is seen from these plots that the quantity $\Delta$I/I shows oscillatory behaviour 
at all the aforementioned locations for the epochs, before and spanning the flare. We 
do observe some spikes in the temporal behaviour of $\Delta$I/I spanning the 
flare for the locations `K2', `K5', and `K6'. However, it could be attributed to the 
effect of the flare as these locations are in the vicinity of H$\alpha$ 
flare-ribbons. We have also estimated the Fourier power spectrum of the variations 
in $\Delta$I/I for all the locations (`K1', `K2', `K3', `K4', `K5', and `K6') 
for the pre-flare phase (03:01-06:01~UT) and 
the flare phase (06:31-09:31~UT) as shown in the Figure~14. It is observed that 
these power spectra show a general enhancement in the oscillatory power during 
the flare as compared to the pre-flare phase for all these locations. 
However, this effect is more for the locations in the vicinity 
of the H$\alpha$ flare-loops (`K2', `K5', and `K6') as compared to the locations 
away from the flare-loops (`K1', `K3', and `K4'). Thus, our results indicate 
that these magnetic jerks could also power the chromospheric oscillations, 
apart from the localized photospheric oscillations. 

These magnetically driven oscillations in the active regions are important for understanding 
the transport of acoustic energy from the photospheric levels to the higher 
solar atmospheric layers.

\begin{acknowledgements}
We are thankful to the anonymous referee for constructive comments and suggestions 
that improved the discussions and presentation of our work in this paper. We acknowledge 
the use of data from HMI instrument on board {\em SDO} spacecraft. The {\em SDO} is a 
mission of NASA aimed to understand the causes of solar variability and its impacts 
on Earth. Our sincere thanks 
to the HMI team for providing the tracked magnetic and velocity data as per our requirement.
This work utilizes the Global Oscillation Network Group (GONG) data obtained by the NSO 
Integrated Synoptic Program (NISP), managed by
the National Solar Observatory, which is operated by AURA, Inc. under a cooperative agreement with
the National Science Foundation. Thanks to GONG++ team for providing the processed H$\alpha$ 
observations. We also acknowledge the use of data from {\em GOES-15} space mission. 
\end{acknowledgements}

\label{lastpage}


\begin{thebibliography}{99}

\bibitem[Burtseva et al.(2015)]{burt15} 
Burtseva, O., Mart{\'{\i}}nez-Oliveros, J.~C., Petrie, G.~J.~D., \& Pevtsov, A.~A.\ 2015, \apj, 806, 173

\bibitem[Domingo et al.(1995)]{domingo95} 
Domingo, V., Fleck, B., \& Poland, A.~I.\ 1995, \solphys, 162, 1

\bibitem[Fisher et al.(1985)]{fisher85} 
Fisher, G.~H., Canfield, R.~C., \& McClymont, A.~N.\ 1985, \apj, 289, 425

\bibitem[Fisher et al.(2012)]{fisher12} 
Fisher, G.~H., Bercik, D.~J., Welsch, B.~T., \& Hudson, H.~S.\ 2012, \solphys, 277, 59

\bibitem[Harvey et al.(1986)]{harvey86} 
Harvey, J.~W.\ 1986, in Small-Scale Magnetic Flux Concentration in the Solar Photosphere, 
ed. W.~Deinzer, M.~Knolker, \& H.~Voight (Gottingen: Vandenhoeck \& Rupprecht), 25 

\bibitem[Harvey \& GONG Instrument Team(1995)]{harvey95} 
Harvey, J., \& GONG Instrument Team 1995, GONG 1994.~Helio- and Astro-Seismology 
from the Earth and Space, 76, 432 

\bibitem[Harvey et al.(1996)]{harvey96} 
Harvey, J.~W., Hill, F., Hubbard, R.~P., et al.\ 1996, Science, 272, 1284 

\bibitem[Harvey et al.(1998)]{harvey98} 
Harvey, J., Tucker, R., \& Britanik, L.\ 1998, Structure and Dynamics of the Interior of 
the Sun and Sun-like Stars, 418, 209

\bibitem[Harvey et al.(2011)]{harvey11} 
Harvey, J.~W., Bolding, J., Clark, R., et al.\ 2011, Bulletin of the American Astronomical Society, 1745 

\bibitem[Hudson et al.(2008)]{hudson08} 
Hudson, H.~S., Fisher, G.~H., 
\& Welsch, B.~T.\ 2008, in ASP Conf. Ser. 383, Subsurface and Atmospheric Influences on Solar Activity, 
ed. R. Howe et al. (San Francisco, CA: ASP), 221

\bibitem[Kaiser et al.(2008)]{kaiser08} 
Kaiser, M.~L., Kucera, T.~A., Davila, J.~M., et al.\ 2008, \ssr, 136, 5 

\bibitem[Kosovichev \& Zharkova (2001)]{koso01}
Kosovichev, A.~G., \& Zharkova, V.~V.\ 2001, \apj, 550, L105

\bibitem[Kosugi et al.(2007)]{kosugi07} 
Kosugi, T., Matsuzaki, K., Sakao, T., et al.\ 2007, \solphys, 243, 3 

\bibitem[Kumar et al.(2011)]{kumar11} 
Kumar, B., Venkatakrishnan, P., Mathur, S., Tiwari, S.~K., \& Garc{\'{\i}}a, R.~A.\ 2011, \apj, 743, 29 

\bibitem[Lin et al.(2002)]{lin02} 
Lin, R.~P., Dennis, B.~R., Hurford, G.~J., et al.\ 2002, \solphys, 210, 3 

\bibitem[Nakajima et al.(1985)]{naka85} 
Nakajima, H., Sekiguchi, H., Sawa, M., Kai, K., \& Kawashima, S.\ 1985, \pasj, 37, 163 

\bibitem[Patterson \& Zirin(1981)]{patt81} 
Patterson, A., \& Zirin, H.\ 1981, \apjl, 243, L99

\bibitem[Patterson(1984)]{patt84} 
Patterson, A.\ 1984, \apj, 280, 884 

\bibitem[Pesnell et al.(2012)]{pes12} 
Pesnell, W.~D., Thompson, B.~J., \& Chamberlin, P.~C.\ 2012, \solphys, 275, 3 

\bibitem[Petrie \& Sudol(2010)]{petrie10} 
Petrie, G.~J.~D., \& Sudol, J.~J.\ 2010, \apj, 724, 1218

\bibitem[Petrie(2014)]{petrie14} 
Petrie, G.~J.~D.\ 2014, \solphys, 289, 3663 

\bibitem[Press et al.(1992)]{press92} 
Press, W.~H., Teukolsky, S.~A., Vetterling, W.~T., \& Flannery, B.~P.\ 1992, 
Cambridge: University Press, |c1992, 2nd ed.,

\bibitem[Qiu \& Gary(2003)]{qiu03} 
Qiu, J., \& Gary, D.~E.\ 2003, \apj, 599, 615 

\bibitem[Scherrer et al.(1995)]{scherrer95} 
Scherrer, P.~H., Bogart, R.~S., Bush, R.~I., et al.\ 1995, \solphys, 162, 129 

\bibitem[Sudol \& Harvey(2005)]{sudol05} 
Sudol, J.~J., \& Harvey, J.~W.\ 2005, \apj, 635, 647

\bibitem[Schou et al.(2012)]{schou12} 
Schou, J., Scherrer, P.~H., Bush, R.~I., et al.\ 2012, \solphys, 275, 229 

\bibitem[Venkatakrishnan et al.(2008)]{venkat08} 
Venkatakrishnan, P., Kumar, B., \& Uddin, W.\ 2008, \mnras, 387, L69 

\bibitem[Wang(1992)]{wang92} 
Wang, H.\ 1992, \solphys, 140, 85  

\bibitem[Wang et al.(1994)]{wang94} 
Wang, H., Ewell, M.~W., Jr., Zirin, H., \& Ai, G.\ 1994, \apj, 424, 436 

\bibitem[Wang et al.(2000)]{wang00} 
Wang, H., Qiu, J., Denker, C., et al.\ 2000, \apj, 542, 1080

\bibitem[Wang et al.(2002)]{wang02} 
Wang, H., Spirock, T.~J., Qiu, J., et al.\ 2002, \apj, 576, 497


\end{thebibliography}
\end{document}